\magnification=1200
\overfullrule=0pt
\baselineskip=20pt
\parskip=0pt
\def\dag{\dagger}
\def\del{\partial}

\def\a{\alpha}     
      
     \def\G{\mit\Gamma}
     
\def\e{\epsilon}   
\def\z{\zeta}

\def\m{\mu}	   
\def\n{\nu}

\def\p{\pi}        
\def\r{\rho}       
\def\s{\sigma}

\def\h{\chi}       
       
\def\w{\omega}     \def\W{\mit\Omega}
\def\br{\langle}
\def\ke{\rangle}
\def\ve{\vert}

{\settabs 5 \columns
\+&&&&UdeM-GPP-TH-98-49\cr}
\bigskip
\centerline{\bf Electromagnetic Interactions in the Quantum Hall Ferromagnet }
\bigskip\bigskip
\centerline{Rashmi Ray$^1$ }
\bigskip

\centerline{Laboratoire de Physique Nucl\'eaire} 
\centerline{ Universit\'e de Montr\'eal}
\centerline{Montr\'eal, Quebec H3C 3J7, Canada}
\bigskip
\centerline{\bf Abstract}
\bigskip
The $\nu =1$ quantum Hall ground state in samples like GaAs is well known to
be ferromagnetic. The global SU(2) spin symmetry of the microscopic action
is broken down to a U(1) symmetry by the ground state. The Goldstone bosons
corresponding to this spontaneous breaking of symmetry are the ferromagnetic
magnons which are neutral spin waves. In addition, there are topologically
nontrivial, electrically charged spin excitations known as spin skyrmions,
which in these samples are the favoured charge carriers. In this letter, we
look into the electromagnetic coupling of these spin excitations. The discrete
symmetries P and T are also broken by the ground state and we find that to the 
leading order, the electromagnetic interaction of the spin waves occurs through
a nonminimal coupling given by a mixed Chern-Simons like term containing both
the electromagnetic potentials and the two Euler angles that specify the coset
SU(2)/U(1) and thereby, the Goldstone bosons.
\vfill
\noindent{$^1$E-mail address: rray@lpshpb.lps.umontreal.ca }

\vfill\eject
\centerline{\bf I. Introduction}
\bigskip
Recently, it has been noticed [1] that the $\nu =1$ quantum Hall state in 
samples like GaAs (where the effective gyromagnetic ratio $g \ll 2$) is
ferromagnetic. In these systems, the $\nu =1$ state breaks the global 
SU(2) symmetry of the microscopic action down to a U(1) symmetry. This is
an instance of Spontaneous Symmetry Breaking (SSB). This state also breaks
the discrete P and T symmetries. The corresponding Goldstone modes are 
the spin waves known as ferromagnetic magnons. They are effectively described
by the two angles needed to describe the coset SU(2)/U(1). The Goldstone
bosons are gapless and neutral. Apart from these, the excitations also include
charged, topologically nontrivial spin excitations known as baby skyrmions [1].
An effective action governing these excitations has been obtained in [1,2,3,4]
. It has also been shown [1] that these skyrmions are the preferred
carriers of electric charge in these systems. 

It would thus be very interesting to investigate the manner in which these
spin excitations couple to electromagnetism. The magnons themselves are 
neutral. Hence we should not expect minimal coupling [5]. In the sequel, we
study the form of the non-minimal coupling by subjecting the microscopic 
fermions to perturbing electromagnetic fields.

\bigskip
\centerline{\bf II. Notation and Formulation}   

\bigskip
The microscopic model for the quantum Hall system in samples like
GaAs may be taken to be:
$$
S = S_1 + S_2 \eqno (2.1)
$$
with
$$
S_1 \equiv \int dt \ \int d\vec{x}\  \psi^{\dag }(\vec x,t)
\bigl[ i\del_{t}-a^{0}-{1\over{2m}}(\vec p - \vec A-\vec a)^2 + \mu \bigr]
\psi(\vec x,t)\eqno (2.2)
$$
$$
S_2 \equiv -{{V_0}\over 2}\int dt \ \int d\vec{x}\  
\bigl( \psi^{\dag } \psi \bigr)^2
\eqno (2.3)
$$
where $\del_{x}A^{y}-\del_{y}A^{x} = -B$, $B$ being the external
magnet field and where $\mu $ is the chemical potential. $a^{\mu }$ are the
perturbative slowly varying fields coupled minimally to the fermions.
$\psi (\vec x,t)$ is a 2-component spinor satisfying the usual
anticommutator 
$$ \{ \psi_{\alpha }(\vec x,t), \psi^{\dag }_{\beta }
(\vec y,t)\} = \delta_{\alpha \beta }\delta (\vec x - \vec y ).
$$

Here we have taken the limit $g\rightarrow 0$ and set the Pauli term
in the action equal to zero. In this limit, the above action has an
exact spin SU(2) symmetry. If restored, a small Pauli term leads to
a soft explicit breaking of this symmetry.

More importantly, we note that in the above action, we have replaced 
the non-local repulsive Coulomb term by a local repulsive 
four-fermi interaction to simplify the analysis.
The long-
distance part can be added on as a perturbation [2].

The partition function of the system is immediately written as
$$
Z = \int D \psi \int D \bar{\psi }\ e^{i(S_1 + S_2)}.\eqno (2.4)
$$

At this point we may use the standard property of the properly
normalised SU(2) generators, 
to re-organise the Coulomb term as [2]
$$
S_2 = V_0\int dt \int d\vec x\ \bigl[ {\vec S}^2 + {1\over 4}{\r }^2
\bigr] \eqno (2.5)
$$
where $\r \equiv \bar \psi \psi $ is the density and
$S^{a} \equiv \bar \psi t^{a} \psi $ is the spin density. Here,
$t^{a}\equiv {{{\sigma }^{a}}\over 2}$, where the $\sigma^{a}$ are the
standard Pauli matrices.

We can now perform the standard Hubbard-Stratonovich transformation.
Upon introducing the auxiliary fields $\vec h \equiv  h \hat{n} $, where
$\hat{n}$ is a unit vector and $\phi $, the partition function is written as
$$
Z = \int D\phi D\vec h\  e^{-{{i}\over {V_0}}\int dtd\vec x\  [{\phi }^2 
+{1\over 4}h^2]}\int D\psi D\bar{\psi }\ e^{i\int dt d\vec x\  \bar{\psi }
[i\del_{t}-a^{0}+\mu +\phi -{1\over{2m}}(\vec p - \vec A -\vec a )^2 
+ {{h}\over 2}
\hat{n}\cdot \vec {\sigma }]\psi }. \eqno (2.6)
$$

Due to SSB, the $\phi $ and the $h$ integrals have the nonzero saddle points
$\phi_0 \simeq {{V_0}\over 2}({{B}\over {2\p }})$ and 
$h^{a}_0 \simeq V_0({{B}\over {2\p }})\delta^{a,z}$. This just means that
the ground state is of uniform density $\r_0 = {{B}\over{2\p }}$ and
uniform magnetisation $\ve \vec h_0 \ve = V_0 \r_0 $, taken to be along
the $\hat z $ direction. This is SSB. 
The fluctuations in $\phi $ and in $h$ are obviously massive while 
Goldstone's theorem guarantees that the fluctuations in $\hat n$, the
Goldstone modes, are gapless.
Upon evaluating the $\phi $ and the $h$ integrals at the saddle points, the
partition function is written as

$$
Z = \int D\hat{n} \int D\psi D\bar{\psi }\ e^{i\int dtd\vec x\  \bar{\psi }
[ i\del_{t}-a^{0}+\mu -{1\over{2m}}(\vec p-\vec A -\vec a)^2 + \z \hat{n}\cdot \s ]
\psi }
\eqno (2.7)
$$

where $\z \equiv {{\r_0 V_0}\over 2}$.

We now introduce the space-time dependent unitary matrix $U \in SU(2)$
such that
$$
U^{\dag }\vec \s \cdot \hat{n}U = \s_{z} . \eqno (2.8)
$$
This naturally induces the SU(2)-valued pure gauge potentials
$$
\W^{a}_{\m }t^{a}\equiv  U^{\dag }i\del_{\m }U .\eqno (2.9)
$$
satisfying
$$
F^{a}_{\m \n }\equiv  \del_{\m }\W^{a}_{\n }-\del_{\n }\W^{a}_{\m }
+ \e^{abc}\W^{b}_{\m }\W^{c}_{\n } = 0. \eqno (2.10)
$$
Let $\h \equiv U^{\dag }\psi $ and $ \bar \h \equiv \bar {\psi } U$.
Thus,
$$
Z = \int D\hat{n} \int D\h D\bar{\h }\ e^{i\int dt d\vec x\  \bar{\h }
[i\del_{t}-a^{0}+\m + \W^{a}_0 t^{a} - {1\over{2m}} 
(-i\del_{i} - A^{i}-a^{i} - \W^{a}_{i}
t^{a})^2 + \z \s_{z}]\h }. \eqno (2.11)
$$

Integrating the fermionic fields $\h , \bar{\h } $ out, we have
$$
Z = \int D\hat{n} \ e^{iS_{eff}}
$$
where
$$
S_{eff}=-i\ {\rm tr} \ln \bigl[ i\del_{t}-a^{0}+\m + \W^{a}_0 t^{a} +\z \s_{z}
-{1\over{2m}}(-i\del_{i}-A^{i}-a^{i}-\W^{a}_{i}t^{a})^2 \bigr]. \eqno (2.12)
$$

We can write the operator-valued argument, $\hat O$ of the above functional 
determinant as
$$
\hat O \equiv  i\del_{t}+\m -h_0 -V \eqno (2.13)
$$

where $h_0$ is the part that can be diagonalised readily and $V$ is the
perturbation.

Here,

$$
h_0 \equiv  {1\over{2m}}(\vec p - \vec A)^2 - \z \s_{z}. \eqno (2.14)
$$

We define $\p^{i} \equiv  -i\del_{i} - A^{i}$. Making the holomorphic and the
anti-holomorphic combinations:
$\p \equiv  \p^{x}-i\p^{y}$ and $\p^{\dag }\equiv  \p^{x}+i\p^{y}$, with
$$
[\p , \p^{\dag }] = 2B \eqno (2.15)
$$
we can rewrite 
$$
h_0 = {1\over{2m}}(\p^{\dag }\p + B) - \z \s_{z}. \eqno (2.16)
$$
The spectrum of this operator is infinitely degenerate (${{B}\over{2\p }}$
states per unit area) and this degeneracy is exposed in terms of the
so-called ``guiding-centre" coordinates 
$X \equiv  x - {1\over{B}}\p^{y}$ and $Y \equiv  y + {1\over{B}}\p^{x}$. 
We form
the combinations 
$Z \equiv  X+iY$ and $\bar Z \equiv  X-iY$ with the commutation relation
$$
[Z,\bar Z]={2\over{B}}. \eqno (2.17)
$$

We see that $X$ (or $Y$) commutes with $h_0$. Thus an eigenbasis for
$h_0$ is chosen to be $\{ \ve n,X,\a \ke \} $
with $n=0,1,2,\dots \infty $, $-\infty \leq X \leq \infty $ and $\a = \pm 1$.
The index $n$ denotes a Landau level (L.L.) and $\a $ denotes the spin 
(whether ``up" or ``down").

In this basis,
$$
h_0 \ve n,X,\a \ke = [(n+{1\over 2})\w_{c} - \z \a ]\ve n,X,\a \ke 
\eqno (2.18)
$$
where $\w_{c}\equiv  {{B}\over {m}}$ is the cyclotron frequency.

In the following, we assume
that $\z \equiv  {{\r_0 V_0}\over 2} \ll \w_{c}$.

The ferromagnetic many-body ground state is constructed out of these
single particle states by filling up all the degenerate states
with $n=0, \a =1$.  

From (2.12) and (2.13), we see that
$$
S_{eff}=-i\ {\rm tr} \ln [i\del_{t}+\m -h_0] + i\ {\rm tr} \sum_{l=1}^{\infty }
{1\over{l}} \bigl( G V \bigr)^{l} \eqno (2.19)
$$
where $[i\del_{t}+\m -h_0]G = I$ and where the logarithm has been expanded. 

Now let us define $\hat p_0$ such that $[\hat t, \hat p_0 ] = -i$
with $\br t \ve \hat p_0 = i\del_{t} \br t \ve $. Let us introduce the
basis $\{ \ve \w \ke \} $ with $\hat p_0 \ve \w \ke = \w \ve \w \ke $.
Furthermore, $ \br \w \ve t \ke = {1\over{\sqrt{2\p }}}\ e^{i\w t}$.

We also introduce the spin projection operators 
$ P_{+}\equiv  {1\over 2}(I + \s_{z})$ and 
$ P_{-}\equiv  {1\over 2}(I - \s_{z})$, which project onto $\a = \pm 1$
respectively.

Now, $\{ \ve n,X,\w \ke \}$ is a basis that diagonalises $ p_0 +\m -h_0 $
and consequently, $G$.

Let
$$
G \ve n,X,\w \ke \equiv {\G }^{(n)}(\w ) \ve n,X,\w \ke . \eqno (2.20)
$$
Given that the ground state is ferromagnetic, we have
$$
\G^{(0)}(\w ) P_{+} = {1\over{\w +\m - {{\w_{c}}\over 2}+\z -i\e }}P_{+}
\eqno (2.21)
$$
$$
\G^{(0)}(\w ) P_{-} = {1\over{\w +\m - {{\w_{c}}\over 2}-\z +i\e }}P_{-}
\eqno (2.22)
$$
and for $n \neq 0$,
$$
\G^{(n)}(\w ) P_{\pm } = {1\over{\w +\m -(n+{1\over 2})\w_{c} \pm \z + i\e }}
P_{\pm } . \eqno (2.23)
$$

Henceforth, we choose $\m = {{\w_{c}}\over 2}-\z $.

The perturbations $V(\hat z \equiv \hat{x} + i\hat{y} , 
\hat{\bar{z}} \equiv \hat{x} - i \hat{y} )$ may be expressed in the
L.L. basis as $V(\hat{Z}-{{i}\over{B}}{\hat \pi }^{\dag },
\hat{\bar{Z}} + {{i}\over{B}}\hat{\pi } )$.
Thus, $V(\hat{x} , \hat{y} )$ may be Taylor expanded around
$\hat Z$ and $\hat {\bar Z} $. If we require a normal ordering prescription
for the $\hat{\pi } , \hat{\pi^{\dagger }}$, this automatically enforces
an anti-normal ordering on $\hat{Z}, \hat{\bar{Z}}$. The notation adopted
here for this anti-normal ordering is
$\sharp \cdots \sharp $.
We further note that $\p ,\p^{\dag } \sim \sqrt{B}$.
Thus this Taylor expansion is also an expansion in inverse powers of $B$.
Therefore, for a large value of $B$ ($\sim 10$ T), the higher derivative terms
should become more and more marginal.

Let us now define
$$
{\cal A}_0 \equiv  a_0 - \W^{a}_0 \eqno (2.24)
$$
and
$$
{\cal A}^{i} \equiv  a^{i} + \W^{a}_{i} . \eqno (2.25)
$$
with $A \equiv  {\cal A}^{x}+i{\cal A}^{y}$, $\bar A \equiv  {\cal A}^{x}-i
{\cal A}^{y}$ and ${\cal B}\equiv \del_{x}{\cal A}^{y} - \del_{y}{\cal A}^{x}$.
Thus the perturbation $V$ is organised as
$$
V=V^{({1\over 2})}+V^{(0)}+V^{(-{1\over 2})}+\cdots \eqno (2.26)
$$
where the ellipses indicate terms subleading in $B$. 
Here,
$$
V^{({1\over 2})}=-{1\over {2m}}\bigl( \sharp A \sharp \p + 
\p^{\dag }\sharp \bar A
\sharp \bigr) \eqno (2.27)
$$
$$
V^{(0)}=\sharp \bigl[{\cal A}^0 - {1\over {2m}}{\cal B} 
+ {1\over{2m}}({\cal A}^{i}
)^2 \bigr] \sharp  \eqno (2.28)
$$
and
$$
V^{-({1\over 2})}={{i}\over{B}}\bigl[ \sharp \del_{\bar Z}{\cal A}^0 \sharp \p 
- \p^{\dag }\sharp \del_{Z}{\cal A}^0 \sharp \bigr]. \eqno (2.29)
$$
\bigskip
\centerline{\bf III. The Effective Action for the Goldstone Modes.}
\bigskip
In this section, we shall compute the effective action given in
(2.19) to $O(1/B)$. 
To the required order, we need to compute
$$
S_{eff}=i\  {\rm tr} \bigl[ GV^{(0)} + {1\over 2}GV^{({1\over 2})}G
V^{({1\over 2})} + GV^{({1\over 2})}GV^{(-{1\over 2})}
+ GV^{({1\over 2})}GV^{({1\over 2})}GV^{(0)} \bigr] \eqno (3.1)
$$
where we have used the cyclic property of the trace.

The simplest term is where there is only one insertion of the
perturbing potential. Namely, we focus on
$$
S^{(1)}\equiv {\rm tr}\ GV . \eqno (3.2)
$$
Upon introducing the basis where $G$ is diagonal,
$$
S^{(1)}=i\ {\rm tr}\  \sum^{\infty }_{n=0}\int dX \int d\w {\G }^{(n)}(\w )
\br n,X,\w \ve V(\hat t)\ve n,X,\w \ke . \eqno (3.3)
$$
Using the fact that
$$
\int {{d\w }\over{2\p }}\ e^{i\w \delta }{\G }^{(n)}(\w )
= iP_{+} \eqno (3.4)
$$
we get
$$
S^{(1)}=-{\rm tr}\  P_{+}\int dt \int dX \br 0,X \ve V^{(0)} \ve 0,X \ke .
\eqno (3.5)
$$
Thus, in real space, we can write, using the explicit expression for
$V^{(0)}$ from (2.28)
$$
S^{(1)}=\int dt \int d\vec x \bigl[ -\r_0 a^0 + {1\over 2}\r_0 {\W }^{z}_0
+ {{\w_{c}}\over{4\p }} + {{\w_{c}}\over {8\p }}(\del_{x}{\W }^{z}_{y}-
\del_{y}{\W }^{z}_{x}) -{{\w_{c}}\over {4\p }}{\rm tr}\ P_{+}({\cal A}^{i})^2
\bigr] \eqno (3.6)
$$
where $\r_0 \equiv {{B}\over{2\p }}$.
Let us now look at the term with two insertions of $V$.
$$
S^{(2)}\equiv {{i}\over 2}{\rm tr} \ G V G V .\eqno (3.7)
$$
Thus,
$$
S^{(2)} = {{i}\over 2}{\rm tr}\ \sum^{\infty }_{n=0} \int d\w \int dX
{\G }^{(n)}(\w )\br n,X,\w \ve V(\hat t) G(\hat p_0)V(\hat t) 
\ve n,X,\w \ke \eqno (3.8)
$$
where the trace is now over the spin indices.

With the same techniques as were used for computing $S^{(1)}$, we may calculate 
$S^{(2)}$: 

$$
S^{(2)}=S^{(2a)}+S^{(2b)}+S^{(2c)} \eqno (3.9)
$$
where
$$
S^{(2a)}=\int dt \int d\vec x \bigl[ {{\w_{c}}\over{4\p }}{\rm tr}\ P_{+}
({\cal A}^{i})^2 -{{\w_{c}}\over{8\p }}(\del_{x}{\W }^{z}_{y}-
\del_{y}{\W }^{z}_{x})-{{\z }\over{8\p }}\bigl( (\vec {{\W }_{i}})^2 - 
({\W }^{z}_{i})^2 \bigr) - {{\z }\over{4\p }}\bigl( \vec{{\W }_{x}}
\times \vec{{\W }_{y}} {\bigr) }^{z}\bigr] \eqno (3.10)
$$
$$
S^{(2b)}+S^{(2c)}=\int dt \int d\vec x \bigl[ 
-{1\over{4\p }}\e^{\m \n \r }a_{\m }\del_{\n }a_{\r } + {1\over{4\p }}
\e^{\m \n \r }a_{\m }\del_{\n }{\W }^{z}_{\r }
+{3\over {16}}\vec{{\W }_0}\cdot (\vec{{\W }_{x}}\times \vec{{\W }_{y}})
\bigr] . \eqno (3.11)
$$

At this point we note a few things about the contributions $S^{(1)}$ and
$S^{(2)}$. There is a gauge non-invariant term in $S^{(1)}$, namely
${\rm tr}\ P_{+}({\cal A}^{i})^2$, which cancels against an equal but opposite
contribution from $S^{(2a)}$. Furthermore, the term $\del_{x}{\W }^{z}_{y}-
\del_{y}{\W }^{z}_{x}$ also cancels between these two. Furthermore, the
external magnetic field breaks the P symmetry and the ferromagnetic
ground state breaks the T symmetry. Thus, one should expect terms that break
these discrete symmetries in the effective action.
There is also
the Hopf term, $\vec{{\W }_0}\cdot (\vec{{\W }_{x}}\times \vec{{\W }_{y}})$
which has been induced in the effective action. However, owing to the
fact that ${\W }^{a}_{\m }$ is a pure gauge potential, the Hopf term receives
corrections from the term in the effective action with three insertions of
the perturbation.

Namely, we have to look at
$$
S^{(3)}\equiv -i\ {\rm tr}\ GV^{({1\over 2})}GV^{({1\over 2})}GV^{(0)}.
\eqno (3.12)
$$
This may be done straightforwardly and yields:
$$
S^{(3)}\simeq \int dt \int d\vec x \bigl[ -{1\over{16\p }}
\vec{{\W }_0}\cdot (\vec{{\W }_{x}}\times \vec{{\W }_{y}})\bigr] .
\eqno (3.13)
$$
Thus combining the terms together, one obtains
$$
\eqalignno{S_{eff} = \int dt \int d\vec x \bigl[ & {1\over 2}\r_0 {\W }^{z}_
{0} - {{\z }\over{8\p }}( (\vec{\W }_{i})^2 - ({\W }^{z}_{i})^2 ) - 
{{\z }\over{4\p }}(\vec{{\W }_{x}} \times \vec{{\W }_{y}})^{z} \cr 
& + {1\over{8\p }}\vec{{\W }_0}\cdot (\vec{{\W }_{x}}\times \vec{{\W }_{y}})
- \r_0a^{0} + {{\w_{c}}\over {4\p }}b - {1\over{4\p }}\e^{\m \n \r }
a_{\m }\del_{\n }a_{\r } \cr 
& + {1\over{4\p }} \e^{\m \n \r }a_{\m }\del_{\n }{\W }^{z}_{\r }\bigr].
& (3.14) \cr 
}
$$
Here, $b\equiv \del_{x}a^{y} - \del_{y}a^{x}$ is the perturbing magnetic
field.

Now, we can express the unitary
matrix $U \in $ SU(2) in terms of the Euler angles, $\theta , \phi ,\h $.
Again, as $ U\s_{z}U^{\dag } = \vec{\s }\cdot {\hat n}$, the unit
vector $\hat n$ is given in terms of the Euler angles as
$$
\hat n = (\sin \theta \cos \phi , \sin \theta \sin \phi , \cos \theta ).
\eqno (3.15)
$$

Thus, we can write the effective action as
$$\eqalignno{
S_{eff}=\int dt \int d\vec x \bigl[ & {{B}\over{4\p }}\cos \theta \del_{t}\phi 
-{{\z }\over{8\p }}(\del_{i}\hat{n})^2 - \z \r_{p} + {1\over{48\p }}
\e^{\m \n \r }\vec{{\W }_{\m }} \cdot (\vec{{\W }_{\n }} \times 
\vec{{\W }_{\r }}) \cr 
& +{1\over {4\p }}\e^{\m \n \r }a_{\m }\del_{\n }{\W }^{z}_{\r }
-\r_0 a^{0} +{{\w_{c}}\over{4\p }}b -{1\over{4\p }}\e^{\m \n \r }
a_{\m }\del_{\n }a_{\r } \bigr] . & (3.16) \cr }
$$
Here, $\r_{p}\equiv {1\over{8\p }}\epsilon^{ij} 
\hat{n}\cdot (\del_{i}\hat{n}\times 
\del_{j}\hat{n})$ is a topological density (the Pontryagin index density).
This means that $\int d\vec x \r_{p} = {\rm integer}$.

Let us look at the various terms in the effective action. The first term,
with a single time derivative is a so called Wess-Zumino term [6].
The second term is the standard kinetic energy
term of the NLSM. Together, the first two terms tell us that the dispersion
relation of the Goldstone bosons are that of ferromagnetic magnons. 
If we set the perturbing electromagnetic field to zero momentarily,
we see that the fourth term, the Hopf term, has the appropriate coefficient
to make the solitons of the NLSM fermionic.  
When the electromagnetic
field is turned on, there is a mixing between the angular degrees of
freedom and the electromagnetic potentials. This is given by the fifth term
in (3.16), which can be rewritten in terms of the Euler angles as
$ -{1\over{4\p }}\sin \theta \e^{\m \n \r }a_{\m }\del_{\n }\theta \del_{\r }
\phi $. This provides a rather elegant expression for the electromagnetic
coupling of the angles parametrising the coset SU(2)/U(1). It is also
gratifying to note that the angle $\h $ has dropped out of the expression,
as it should, since only two angles are needed to describe the coset.
It is well known that skyrmions that exist as topological excitations
in the system are characterised by their winding number 
$N_{p}\equiv \int d\vec x \r_{p}$. the same winding number also gives
the electrical charge of the skyrmion ($e=1$). Thus it is natural that
the response of the system to an electrostatic potential $a^0$ should
be given by a term $\int dt \int d\vec x \r_{p}a^{0} $ in the effective
action. 
In (3.16), the electromagnetic interaction (the fifth term) can be written as
$$
S^{\rm em}_{eff} = -\int dt \int d\vec x \bigl[
\r_{p}a^{0} + {1\over{4\p }}\{ a^{x}(\del_{y}{\W }^{z}_0 - \del_0{\W }^{z}_{y})
+ a^{y}(\del_0{\W }^{z}_{x} - \del_{x}{\W }^{z}_0 )\} \bigr] .
\eqno (3.17)
$$
The first term is as expected. The second term, within parentheses does not
meet with our naive expectations. In terms of the perturbing electromagnetic
fields, this term can be expressed as
$$
S^{\rm em}_{eff}=-{1\over{4\p }}\bigl[ {\W }^{z}_0 b + {\W }^{z}_{x}e^{y}-
{\W }^{z}_{x}e^{x} \bigr] \eqno (3.18)
$$
where
$$
b\equiv \del_{x}a^{y}-\del_{y}a^{x}
$$
and
$$
e^{i}\equiv -(\del_0 a^{i} + \del_{i}a^{0}).
$$
What is quite remarkable is that this second term does not depend on the
details of the interaction $V_{0}$.

From the effective action, we can readily compute the mean electromagnetic 
currents in the spin-textured (excited) state. 

Thus,
$$
\br j_{0} \ke = \r_{0} + \r_{p} -{1\over{2\p }}b
$$
$$
\br j_{x} \ke = {1\over{2\p }}e^{y} + {1\over{4\p }}(\del_{y}{\W }^{z}_{0}
-\del_{0}{\W }^{z}_{y})
$$
$$
\br j_{y} \ke = -{1\over {2\p }}e^{x} + {1\over{4\p }}(\del_{0}{\W }^{z}_{x}
- \del_{x}{\W }^{z}_{0}) . \eqno (3.19)
$$

This shows explicitly that the density in the excited state changes from
its ground state value by an amount which is a topological index density.

Given the effective action, one can also compute the mean magnetisation in the
excited state.
It is given by
$$
M^{a}(\vec x,t) \equiv \br \bar{\h } t^{a}\h \ke = {{\del {\cal L}_{eff}}\over 
{\del {\W }^{a}_{0}}} . \eqno (3.20)
$$

For instance the $z$ component of the magnetisation changes from its
value of ${1\over 2}\r_{0}$ in the ground state. It is given by
$$
M^{z}(\vec x,t)= {1\over 2}\r_{0} + {1\over 2}\r_{p} - {1\over{4\p }}b
\eqno (3.21)
$$
which is precisely equal to half the value of the mean density in the
excited state.
\bigskip
\centerline{\bf V.  Conclusions}
\bigskip
In this letter, we have derived some new results 
on the electromagnetic interaction of the skyrmions in quantum Hall 
ferromagnets.

We have shown how a non-linear sigma model, describing the
spin excitations of the electrons, emerges simply in terms of the angular
variables describing the deviation of the magnetisation vector from its
ferromagnetic ground state orientation. We have shown further that a Hopf
term also emerges in terms of these same angular variables, with a coefficient
that is appropriate for turning the skyrmionic excitations in the system into
fermions.

Furthermore, we
have also investigated the obviously non-minimal electromagnetic coupling
of the spin excitations in the system. To the leading order, we get a
gauge invariant ``Chern-Simons like" term which couples the angular
variables describing the spin excitations, to externally applied 
electromagnetic fields. This term clearly shows how the electromagnetic
currents in the system are affected by the spin excitations. This, we believe
is a new result. 

Within our approach, the U(1) valued electromagnetic potentials are treated
on the same footing as the SU(2) valued potentials describing the spin
dynamics. Since P and T are violated in the system, the emergence of
``Chern-Simons like' terms is only to be expected. In fact, three such terms
are obtained: A pure electromagnetic CS term, a Hopf term purely in terms of 
the angular degrees of freedom and the term described in the previous paragraph
which mixes the two.

If we set all angular excitations to zero, thereby freezing the spin degree
of freedom, we obtain the effective electromagnetic interaction of planar
polarised electrons, as has been described, for instance, in [7,8]. On the
other hand, upon setting the perturbative electromagnetic fields to zero, we
obtain the magnon effective action of [1,2]. Thus our work also provides an
unifying treatment of these two different systems.

The details of the calculations leading to the above results will be presented
elsewhere.
\bigskip
\centerline{\bf Acknowledgements}
\bigskip
I wish to acknowledge J. Soto for early discussions on the subject
and for a fruitful collaboration on a previous article on a similar topic.
T.H. Hansson must be thanked for suggesting that the spin collective
modes might be extracted through a unitary transformation.
I am indebted to R. Mackenzie and M. Paranjape for sharing their insights
with me and B. Sakita and V.P. Nair for their encouragement.
The work is partially supported by the N.S.E.R.C of Canada and the 
F.C.A.R of Quebec.
\vfill\eject
\centerline{\bf References}
\bigskip
\item{[1]} S.L. Sondhi et. al., Phys. Rev. {\bf B 47}, 16419, (1993).
\item{[2]} K. Moon et. al., Phys. Rev. {\bf B 51}, 5138, (1995).
\item{[3]} W. Apel \& Yu.A. Bychkov, Phys. rev. Lett. {\bf 78}, 2188, (1997).
\item{[4]} R. Ray \& J. Soto, cond-mat/9708067.
\item{[5]} J.M. Roman \& J. Soto, cond-mat/9709298.
\item{[6]} E. Fradkin, {\it Field Theories of Condensed Matter Systems},
(Addison-Wesley Publishing Company).
\item{[7]} R. Ray \& B. Sakita, Ann. Phys. {\bf 230}, 131, (1994).
\item{[8]} R. Ray \& J. Soto, Phys. Rev. {\bf B 54}, 10709, (1995).
\end